\documentclass{webofc}
\usepackage[varg]{txfonts}   % Web of Conferences font

\usepackage{subcaption}

\begin{document}

\title{Non-perturbative insights into the spectral properties of QCD at finite temperature}

\author{\firstname{Peter} \lastname{Lowdon}\inst{1}\fnsep\thanks{\email{lowdon@itp.uni-frankfurt.de}} \and
        \firstname{Owe} \lastname{Philipsen}\inst{1}\fnsep\thanks{\email{philipsen@itp.uni-frankfurt.de}} 
}

\institute{Institut f\"{u}r Theoretische Physik, Goethe-Universit\"{a}t,\\  Max-von-Laue-Str. 1,  60438 Frankfurt am Main, Germany}

\abstract{In quantum field theories at finite temperature spectral functions describe how particle systems behave in the presence of a thermal medium. Although data from lattice simulations can in principle be used to determine spectral function characteristics, existing methods rely on the extraction of these quantities from temporal correlators, which requires one to circumvent an ill-posed inverse problem. In these proceedings we report on a recent approach that instead utilises the non-perturbative constraints imposed by field locality to extract spectral function information directly from spatial correlators. In particular, we focus on the application of this approach to lattice QCD data of the spatial pseudo-scalar meson correlator in the temperature range $220-960 \, \text{MeV}$, and outline why this data supports the conclusion that there exists a distinct pion state above the chiral pseudo-critical temperature $T_{\!\text{pc}}$.}

\maketitle

\section{Introduction}
\label{intro}

The dynamical properties of quantum field theories (QFTs) at finite temperature are entirely encoded in the correlation functions of the corresponding quantum fields, and hence the determination of these quantities is essential for describing the effects experienced by particles as they move through a thermal medium. Lattice simulations provide a powerful approach with which one can directly compute the form of correlation functions, and have led to significant advancements in the understanding of finite-temperature phenomena, particularly in QCD. By construction, this method works in imaginary time $\tau$, and for understanding the spectral properties of a theory the quantities of particular interest are the two-point functions of gauge-invariant operators $O_{\Gamma}(\tau,\mathbf{x})$
\begin{align}
C_{\Gamma}(\tau,\mathbf{x}) = \langle O_{\Gamma}(\tau,\mathbf{x})\,O_{\Gamma}^{\dagger}(0,\mathbf{0})\rangle_{T},
\label{corrT}
\end{align}
where $\Gamma$ denotes a set of quantum numbers, and the expectation value is over a thermal ensemble at temperature $T=1/\beta$. Upon taking the spatial Fourier transform of Eq.~\eqref{corrT} it follows from the assumption of thermal equilibrium~\cite{Kapusta:2006pm,Bellac:2011kqa} that 
\begin{align}
\widetilde{C}_{\Gamma}(\tau,\mathbf{p}) = \int_{0}^{\infty} \frac{d\omega}{2\pi} \frac{\cosh\left[\left(\tfrac{\beta}{2}-|\tau| \right)\omega\right] }{\sinh\left(\tfrac{\beta}{2}\omega\right)} \,\rho_{\Gamma}(\omega,\mathbf{p}), 
\label{corrTp}
\end{align}
where the spectral function $\rho_{\Gamma}(\omega,\mathbf{p})$ corresponds to the thermal commutator of the operator $O_{\Gamma}(\tau,\mathbf{x})$. Equation~\eqref{corrTp} is highly significant because it demonstrates that physical characteristics, namely the information about particle excitations in the quantum number channel $\Gamma$ contained in  $\rho_{\Gamma}(\omega,\mathbf{p})$, can be extracted from imaginary-time data. In the specific case $\mathbf{p}=0$ the resulting correlation function $\widetilde{C}_{\Gamma}(\tau)$ coincides with the spatial integral over $C_{\Gamma}(\tau,\mathbf{x})$, and is called the \textit{temporal} correlator. If one instead fixes a spatial direction $x_{3}$, and integrates $C_{\Gamma}(\tau,\mathbf{x})$ over $\tau$ and the remaining spatial directions, one obtains the \textit{spatial} correlator $C_{\Gamma}(x_{3})$, which similarly to the temporal correlator can also be related to the real-time spectral function, in this case via 
\begin{align}
C_{\Gamma}(x_{3}) = \int_{-\infty}^{\infty}  \frac{dp_{3}}{2\pi}e^{i p_{3} x_{3}} \int_{0}^{\infty}  \frac{d\omega}{\pi \omega}  \ \rho_{\Gamma}(\omega,p_{1}=p_{2}=0,p_{3}). \label{C_rho}
\end{align}
Since the dependence on $\rho_{\Gamma}(\omega,\mathbf{p})$ in Eqs.~\eqref{corrTp} and~\eqref{C_rho} is different, lattice simulations of temporal and spatial correlators, either continuum extrapolated or evaluated at the same lattice spacing, provide distinct ways of probing the structure of spectral functions. \\

\noindent
In the literature, significant focus has been given to extracting spectral functions from temporal correlators. A fundamental issue is that this requires one to overcome an ill-posed inverse problem, namely that discrete temporal correlator data and the structure of Eq.~\eqref{corrTp} are not sufficient to uniquely reconstruct the form of $\rho_{\Gamma}(\omega,\mathbf{p})$. For this reason, most strategies for obtaining spectral functions require intricate statistical methods combined with additional input, either from perturbative calculations or phenomenological modelling~\cite{Asakawa:2000tr,Meyer:2011gj}. In the case of spatial correlators, particular progress has been made in establishing their properties in QCD in recent years~\cite{Laermann:2001vg,Wissel:2005pb,Cheng:2010fe,Banerjee:2011yd,Karsch:2012na,Brandt:2014uda,Bazavov:2014cta,DallaBrida:2021ddx}. At large distances it is expected that these correlators decay exponentially 
\begin{align}
C_{\Gamma}(x_{3}) \sim  e^{- \, m_{\Gamma}^{\text{scr}} x_{3}},
\end{align}
where $m_{\Gamma}^{\text{scr}}$ is referred to as the \textit{screening mass}. $m_{\Gamma}^{\text{scr}}$ is temperature dependent and provides a measure of the in-medium modifications experienced by the lowest-energy state created by $O_{\Gamma}$. A significant advantage of spatial correlators is that they can be calculated at arbitrarily large distances, whereas the extent of temporal correlators is limited by the inverse temperature $\beta$ due to their periodicity. In this sense, spatial correlators are more sensitive to the properties of the thermal system. \\      

\noindent
In these proceedings we provide an overview of a novel approach developed in Ref.~\cite{Lowdon:2022xcl} for extracting spectral function information from spatial correlators. We will outline the theoretical foundations upon which this approach is based, and its application to lattice QCD data of the spatial pseudo-scalar meson correlator at different temperatures above the chiral pseudo-critical temperature $T_{\!\text{pc}}$.

\section{Locality constraints}
\label{local}

\subsection{Thermal spectral representation}
\label{thermal}

In order to fully understand the characteristics of QFTs one necessarily requires a framework that is independent of the specific interaction regime of the system. Over the years, axiomatic formulations of local QFT have been developed in order to construct such a non-perturbative framework, and their applications have led to numerous foundational insights, including the relationship between spin and statistics, and the rigorous connection of Minkowski and Euclidean QFTs~\cite{Streater:1989vi,Haag:1992hx,Bogolyubov:1990kw}. However, these insights have mostly been restricted to systems with zero temperature, and it largely remains an open question as to how they should generalise when $T>0$. In Refs.~\cite{Bros:1992ey,Bros:1995he,Bros:1998ua,Bros:1996mw,Bros:2001zs} significant steps were taken to understand how this generalisation might work for simple classes of QFTs involving scalar fields $\phi(x)$. The authors established that together with the well-known conditions brought about by the assumption of thermal equilibrium~\cite{Kapusta:2006pm,Bellac:2011kqa}, the \textit{locality} of the fields\footnote{By locality we mean: $\left[\phi(x),\phi(y)\right]=0$ for $(x-y)^{2}<0$, which corresponds to the physical assumption that all measurements respect causality.} imposes significant constraints. In the case of the spectral function, this implies that it satisfies the following representation
\begin{align}
\rho(\omega,\mathbf{p}) = \int_{0}^{\infty} \!\! ds \int \! \frac{d^{3}u}{(2\pi)^{2}} \ \epsilon(\omega) \ \delta\!\left(\omega^{2} - (\mathbf{p}-\mathbf{u})^{2} - s \right)\widetilde{D}_{\beta}(\mathbf{u},s).
\label{spectral_rep}
\end{align}
Equation~\eqref{spectral_rep} is highly significant because it describes in general how the spectral properties of a QFT are affected by temperature. In fact, Eq.~\eqref{spectral_rep} represents the $T>0$ generalisation of the well-known K\"{a}ll\'{e}n-Lehmann spectral representation, a cornerstone result of local QFT~\cite{Kallen:1952zz,Lehmann:1954xi}. One can see from Eq.~\eqref{spectral_rep} that the temperature dependence is entirely captured by the \textit{thermal spectral density} $\widetilde{D}_{\beta}(\mathbf{u},s)$, and so determining the properties of this quantity is of central importance for understanding the in-medium modification of scalar particles. \\ 
 
\noindent
In Ref.~\cite{Bros:1992ey} the authors proposed that if a theory contains a stable particle state of mass $m$ at $T=0$, and no phase transition is met for $T>0$, then the singular structure of $\widetilde{D}_{\beta}(\mathbf{u},s)$ in the variable $s$ is preserved relative to the vacuum theory, and hence one can write 
\begin{align}
\widetilde{D}_{\beta}(\mathbf{u},s)= \widetilde{D}_{m,\beta}(\mathbf{u})\, \delta(s-m^{2}) + \widetilde{D}_{c, \beta}(\mathbf{u},s),
\label{decomp}
\end{align} 
where $\widetilde{D}_{c, \beta}(\mathbf{u},s)$ is continuous in $s$. There are several reasons for why the decomposition in Eq.~\eqref{decomp} is natural for describing the presence of particle states at finite temperature:
\vspace{2mm}
\begin{itemize}
\item When $T>0$ the \textit{damping factor} $\widetilde{D}_{m,\beta}(\mathbf{u})$ is non-trivial, and due to the structure of Eq.~\eqref{spectral_rep} this causes the $T=0$ peak in $\rho(\omega,\mathbf{p})$ to become broadened outside of the mass shell $p^{2}=m^{2}$. Moreover, the properties of $\widetilde{D}_{m,\beta}(\mathbf{u})$, and hence the precise nature of this thermal broadening, has been shown to depend on the underlying dynamics~\cite{Bros:2001zs}. 

\item The structure of the discrete term implies that the dissipative effects experienced by particle states due to their interactions with the thermal medium is velocity dependent, as one would expect~\cite{Bros:1996mw,Bros:2003zs}.    

\item The decomposition in Eq.~\eqref{decomp} can be generalised to describe intrinsically unstable states by replacing $\delta(s-m^{2})$ with a suitable resonance-type function in $s$, such as a relativistic Breit-Wigner. The factorisation of the $(\mathbf{u},\beta)$ and $s$ dependence then ensures that the representation distinguishes between particle decays of different physical origin; those due to in-medium effects controlled by $(\mathbf{u},\beta)$, and those brought about by the instability of the underlying $T=0$ particle states. 

\end{itemize}
\vspace{2mm}
\noindent
In Ref.~\cite{Bros:2001zs} the specific structure of damping factors were explored in different models, and more recently in Refs.~\cite{Lowdon:2021ehf,Lowdon:2022keu,Lowdon:2022ird} it was shown that these quantities can be used to perform non-perturbative analytic calculations of in-medium observables such as the shear viscosity. 

\subsection{The spatial correlator}
\label{spatial}

In light of the thermal spectral representation in Eq.~\eqref{spectral_rep} it was shown in Ref.~\cite{Lowdon:2022xcl} that the spatial correlator of a scalar operator can be written in the following general manner
\begin{align}
C(x_{3}) = \frac{1}{2}\int_{0}^{\infty} \! ds \int^{\infty}_{|x_{3}|} \! dR \ e^{-R\sqrt{s}} D_{\beta}(R,s),
\label{C_int}
\end{align}
where $D_{\beta}(R,s)$ is the inverse Fourier transform of $\widetilde{D}_{\beta}(\mathbf{u},s)$, and depends only on $R=|\mathbf{x}|$ due to the isotropy of the thermal medium. Equation~\eqref{C_int} demonstrates precisely how the spectral properties of a given theory affects the behaviour of the spatial correlator. Assuming that the decomposition in Eq.~\eqref{decomp} holds, it follows from the structure of Eq.~\eqref{C_int} that if the continuous component is non-vanishing for $s \geq s_{c} \gg m^{2}$, or is suppressed relative to the discrete particle contribution, the corresponding damping factor $D_{m,\beta}(R)$ will dominate the behaviour of the spatial correlator
\begin{align}
C(x_{3}) \approx  \frac{1}{2}\int^{\infty}_{|x_{3}|} \! dR \  e^{-m R} D_{m,\beta}(R),
\label{C_decomp_dom}
\end{align}
and this domination will be especially pronounced at large distances. In QCD, hadronic correlators generally receive contributions from multiple states, and in this case the previous arguments can be suitably extended~\cite{Lowdon:2022xcl}. The ultimate conclusion is that the lowest energy states will have the most influence on $C(x_{3})$ at large distances, and the damping factors associated with these states can be directly extracted from the large-$x_{3}$ behaviour of $C(x_{3})$. This is highly significant, since once the explicit form of the damping factors are known, Eq.~\eqref{spectral_rep} can be used to calculate their \textit{exact} analytic contributions to the full spectral function $\rho(\omega,\mathbf{p})$.

\section{Pseudo-scalar meson correlators in QCD}

\subsection{Spatial correlator analysis}

Given the non-perturbative constraints outlined in Sec.~\ref{local}, one can apply these relations in order to extract spectral function information from lattice spatial correlator data. Here we provide an overview of the study performed in Ref.~\cite{Lowdon:2022xcl} which focussed on the analysis of QCD spatial pseudo-scalar meson correlators, defined by
\begin{align}
C_{\text{PS}}(x_{3}) =  \int^{\infty}_{-\infty} \! dx_{1} \int^{\infty}_{-\infty} \! dx_{2} \int_{-\frac{\beta}{2}}^{\frac{\beta}{2}} d\tau \, \langle \Omega_{\beta}|O_{\text{PS}}^{a}(\tau, \mathbf{x})O_{\text{PS}}^{a \, \dagger}(0)|\Omega_{\beta}\rangle,
\label{PS_C_def}
\end{align} 
where $O_{\text{PS}}^{a}= \overline{\psi}\gamma_{5}\frac{\tau^{a}}{2}\psi$, with $\psi(x)=(u(x),d(x))$ isospin doublets, and $\tau^{a}$ the Pauli spin matrices. This study analysed the lattice QCD data from Ref.~\cite{Rohrhofer:2019qwq} which used two mass-degenerate flavours of domain wall fermions at temperatures $T=$ 220, 320, 380, 480, 660, and 960~MeV. More details regarding the precise lattice setup can be found in Ref.~\cite{Rohrhofer:2019qwq}. Following the strategy in Sec.~\ref{spatial} we performed fits to these data at each temperature value, and found it to be well-described by the two-exponential ansatz
\begin{align}
C_{\text{PS}}(x_{3}) = A e^{-m_{\pi}^{\text{scr}} x_{3}} + B e^{-m_{\pi^{*}}^{\text{scr}} x_{3}}.
\label{C_PS}
\end{align}
As discussed in Sec.~\ref{intro}, the exponents have the interpretation of screening masses for the two lightest states generated by the pseudo-scalar meson operator $O_{\text{PS}}^{a}$, which in this case are the pion $\pi$ and its first radial excitation $\pi^{*}$. Given this pure exponential behaviour it was shown in Ref.~\cite{Lowdon:2022xcl} from the spectral constraints outlined in Sec.~\ref{spatial} that Eq.~\eqref{C_PS} is consistent with the pseudo-scalar spectral function $\rho_{\text{PS}}(\omega,\mathbf{p})$ containing the thermal spectral density component    
\begin{align}
D_{\beta}(\mathbf{x},s)= \sum_{i=\pi,\pi^{*}} \alpha_{i}e^{-\gamma_{i}|\mathbf{x}|}\, \delta(s-m_{i}^{2}), 
\label{D_PS}
\end{align}
where the temperature-dependent damping factor exponents $\gamma_{i}$ are related to the screening masses $m_{i}^{\text{scr}}$ and particle vacuum masses $m_{i}$ as follows
\begin{align}
\gamma_{i} = m_{i}^{\text{scr}} - m_{i}, \quad\quad i=\pi, \, \pi^{*}. 
\label{m_screen}
\end{align}
Now that the explicit functional form of $D_{\beta}(\mathbf{x},s)$ is known one can apply Eq.~\eqref{spectral_rep} to establish precisely how the $\pi$ and $\pi^{*}$ states contribute to $\rho_{\text{PS}}(\omega,\mathbf{p})$. Doing so, one finds
\begin{align}
\rho_{\text{PS}}(\omega,\mathbf{p})&= \epsilon(\omega) \left[ \theta(\omega^{2}-m_{\pi}^{2}) \,  \frac{4 \, \alpha_{\pi} \gamma_{\pi}  \sqrt{\omega^{2}-m_{\pi}^{2}}}{(|\mathbf{p}|^{2}+m_{\pi}^{2}-\omega^{2})^{2} + 2(|\mathbf{p}|^{2}-m_{\pi}^{2}+\omega^{2})\gamma_{\pi}^{2}+\gamma_{\pi}^{4} }            \right. \nonumber \\
& \quad\quad\quad\quad + \left. \theta(\omega^{2}-m_{\pi^{*}}^{2}) \,  \frac{4 \, \alpha_{\pi^{*}} \gamma_{\pi^{*}}  \sqrt{\omega^{2}-m_{\pi^{*}}^{2}}}{(|\mathbf{p}|^{2}+m_{\pi^{*}}^{2}-\omega^{2})^{2} + 2(|\mathbf{p}|^{2}-m_{\pi^{*}}^{2}+\omega^{2})\gamma_{\pi^{*}}^{2}+\gamma_{\pi^{*}}^{4} }     \right].
\label{commutator_PS}
\end{align}    
In order to interpret the physical implications of Eq.~\eqref{commutator_PS} it is simpler to consider the corresponding zero-momentum expression $\rho_{\text{PS}}(\omega):=\rho_{\text{PS}}(\omega,\mathbf{p}=0)$, which has the form
\begin{align}                                                   
\rho_{\text{PS}}(\omega)=   \epsilon(\omega) \left[ \theta(\omega^{2}-m_{\pi}^{2}) \, \frac{4\, \alpha_{\pi} \,  \gamma_{\pi} \sqrt{\omega^{2}-m_{\pi}^{2}}}{(\omega^{2}-m_{\pi}^{2}+\gamma_{\pi}^{2})^{2}} +\theta(\omega^{2}-m_{\pi^{*}}^{2}) \, \frac{4\, \alpha_{\pi^{*}} \,  \gamma_{\pi^{*}} \sqrt{\omega^{2}-m_{\pi^{*}}^{2}}}{(\omega^{2}-m_{\pi^{*}}^{2}+\gamma_{\pi^{*}}^{2})^{2}}   \right].
\label{spectral_pzero}
\end{align}  
Equation~\eqref{spectral_pzero} possesses several distinct characteristics, including the existence of energy thresholds at $\omega= m_{\pi},m_{\pi^{*}}$, and the presence of peaked contributions for each of the particle components, with the locations of the peaks given by
\begin{align}
\omega_{i}^{\text{peak}} = \sqrt{m_{i}^{2} + \tfrac{1}{3}\gamma_{i}^{2}}, \quad\quad  i=\pi, \, \pi^{*}.
\label{peak}
\end{align}
The temperature-dependent parameters $\gamma_{i}$ have the interpretation of thermal width parameters since they control both the relative size and location of the spectral function peaks via Eq.~\eqref{peak}. In the $T\rightarrow 0$ limit $\gamma_{i}$ become increasingly smaller, and Eq.~\eqref{commutator_PS} approaches the zero-temperature form: $2\pi  \epsilon(\omega) \sum_{i=\pi,\pi^{*}} \delta(p^{2}-m_{i}^{2})$, as expected. \\

\begin{figure}[t]
\begin{subfigure}{0.49\linewidth}
\centering
\includegraphics[width=1\columnwidth]{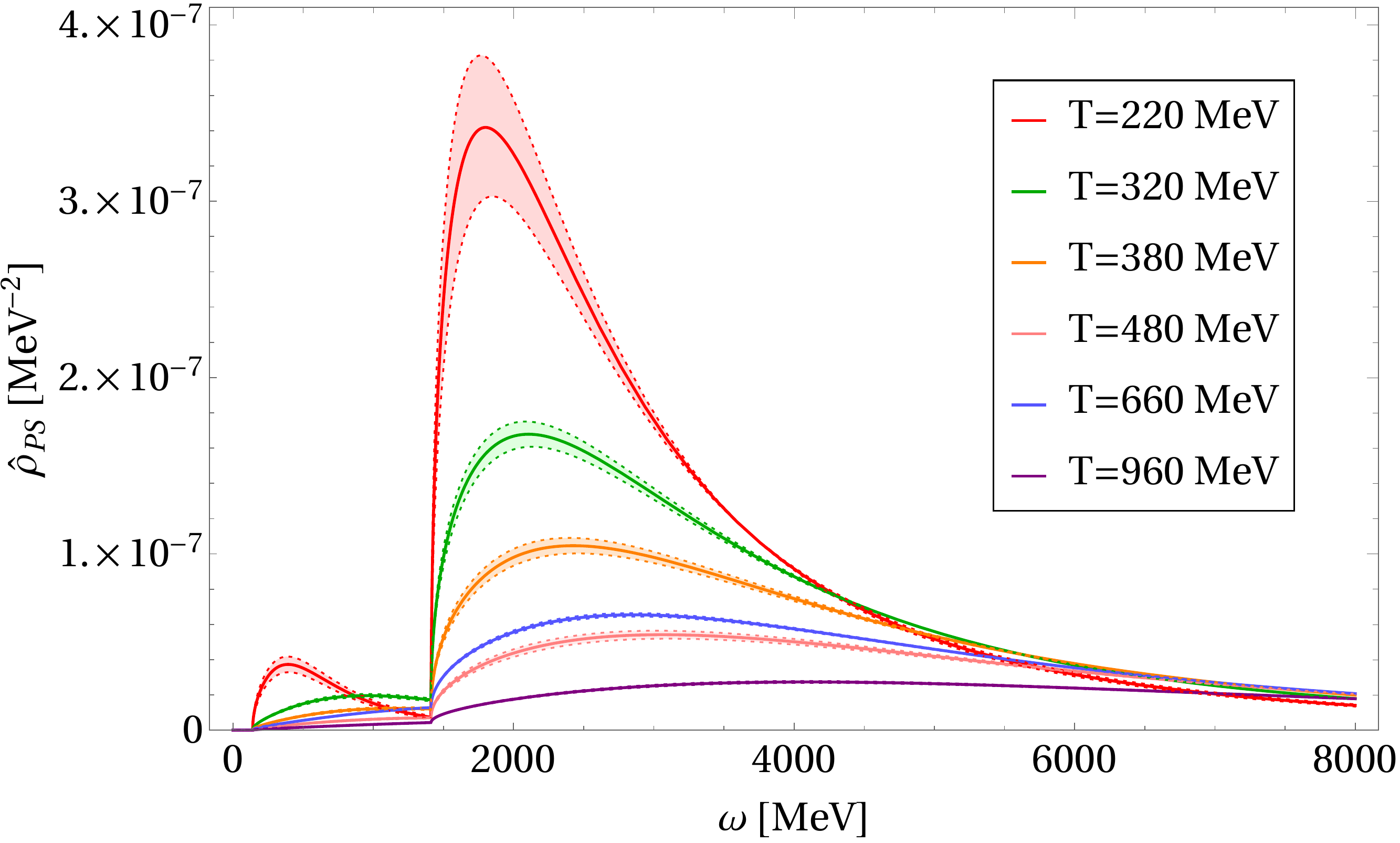}
\end{subfigure} 
\begin{subfigure}{0.495\linewidth}
\centering
\includegraphics[width=1\columnwidth]{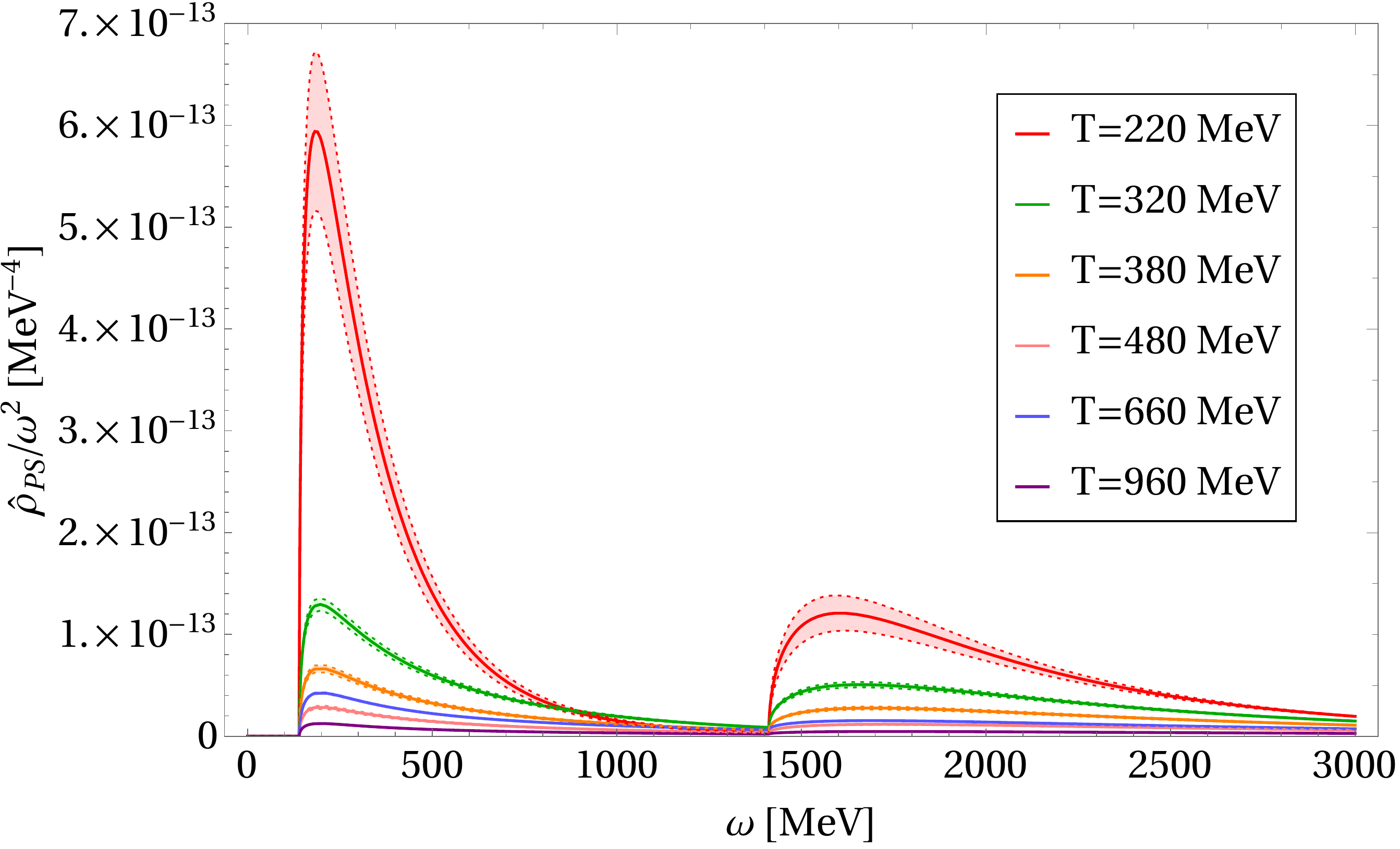}
\end{subfigure}
\caption{The unit-normalised pseudo-scalar spectral function $\hat{\rho}_{\text{PS}}(\omega)$ (left), and rescaled expression $\hat{\rho}_{\text{PS}}(\omega)/\omega^{2}$ (right), at different temperatures for $m_{\pi}=140$~MeV. The coloured bands indicate the 1$\sigma$ uncertainties.}
\label{spectral_plot} 
\end{figure}

\noindent
Using the values of the parameters extracted from the fit ansatz in Eq.~\eqref{C_PS}, together with Eq.~\eqref{spectral_pzero}, we were able to evaluate the explicit form of the spectral function at each of the available temperatures. This is displayed in Fig.~\ref{spectral_plot}. The left plot shows the unit-normalised spectral function $\hat{\rho}_{\text{PS}}(\omega)$ with its respective 1$\sigma$ error band for $m_{\pi}=140$~MeV, and in the right plot is the corresponding rescaled expression $\hat{\rho}_{\text{PS}}(\omega)/\omega^{2}$. The most prominent feature in Fig.~\ref{spectral_plot} (left) is that the pion component of the spectral function gives a pronounced peak at the lowest temperature $T=220$~MeV, which corresponds to approximately $1.2 \, T_{\!\text{pc}}$. As the temperature rises this peak eventually disappears due to increased thermal broadening, and because the peak location $\omega_{\pi}^{\text{peak}}$ begins to exceed the $\pi^{*}$ threshold at $\omega=m_{\pi^{*}}$. Although the $\pi^{*}$ features a distinct peak up to around $T=660$~MeV it is expected that the inclusion of higher-excited pion states, which are not captured by the lattice data, would modify this picture. \\

\noindent
Overall, Fig.~\ref{spectral_plot} supports the physical picture that the pion and its excitations are sequentially melted, and that this process happens gradually and at temperatures significantly larger than the effective chiral symmetry restoration scale $T_{\!\text{pc}}$. This picture is also consistent with the expectations of chiral spin symmetry, which predicts the existence of hadron-like states where chiral symmetry is effectively restored but quarks are still bound~\cite{Rohrhofer:2019qwq,Rohrhofer:2019qal,Glozman:2022lda}, and has been shown to be approximately realised in the temperature range $T_{\!\text{pc} }  \lesssim T  \lesssim 3 T_{\!\text{pc}}$ on the same lattice configurations~\cite{Rohrhofer:2019qwq,Rohrhofer:2019qal}. 

\subsection{Temporal correlator comparison}

\begin{figure}[h] 
\centering
\includegraphics[width=0.6\columnwidth]{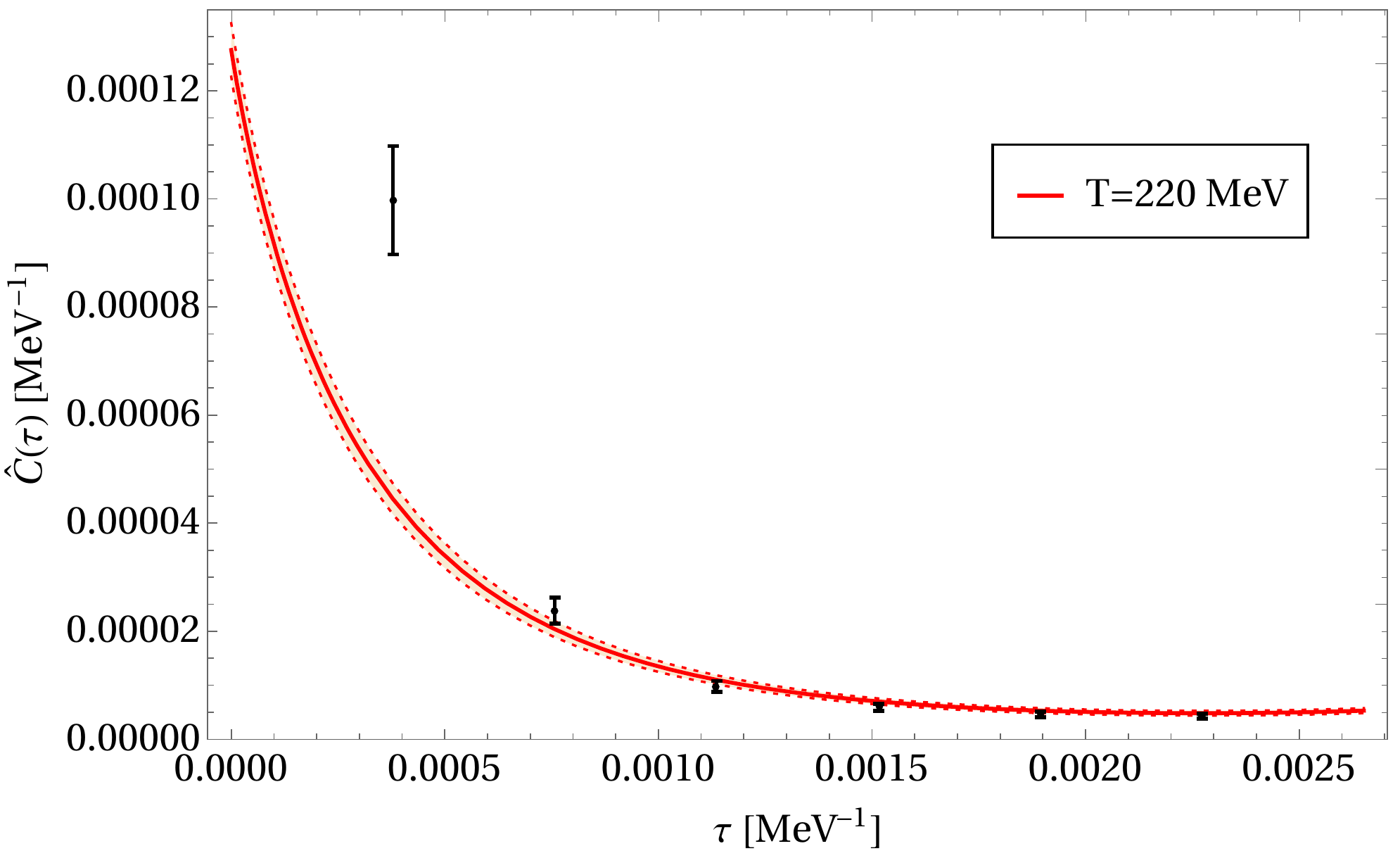}
\caption{The $\hat{C}_{\text{PS}}(\tau)$ prediction against the rescaled lattice data (black points) for $T=220$~MeV. The error bars and coloured band indicate the estimated combined uncertainties.} 
\label{T_corr_plot} 
\end{figure}

As outlined in Sec.~\ref{intro}, spatial and temporal correlators have very different dependencies on the spectral function. This therefore presents a straightforward and non-perturbative test of the consistency of any extracted spectral function: \textit{the spectral function determined from spatial correlators must predict the corresponding temporal ones, and vice versa.} In Ref.~\cite{Rohrhofer:2019qal} the authors generated temporal correlator data at $T=220$~MeV with the same lattice setup as that used to calculate the spatial correlator in Ref.~\cite{Rohrhofer:2019qwq}. By combining Eq.~\eqref{corrTp} (at $\mathbf{p}=0$) with Eq.~\eqref{spectral_pzero} one can therefore use the extracted spectral function at $T=220$~MeV to predict the form of the temporal correlator and compare with the data of Ref.~\cite{Rohrhofer:2019qal}. Since the spectral functions in Fig.~\ref{spectral_plot} are unit normalised, in order to be consistent, the corresponding temporal correlator prediction $\hat{C}_{\text{PS}}(\tau)$ must be compared with the lattice data of Ref.~\cite{Rohrhofer:2019qwq} which is rescaled with the same normalisation. The details of this rescaling procedure are described in Ref.~\cite{Lowdon:2022xcl}. The comparison of the $\hat{C}_{\text{PS}}(\tau)$ prediction and rescaled lattice data at $T=220$~MeV are displayed in Fig.~\ref{T_corr_plot}. \\

\noindent
One can see from Fig.~\ref{T_corr_plot} that the $\hat{C}_{\text{PS}}(\tau)$ prediction provides a good description of the lattice data at large values of $\tau$. As $\tau$ becomes smaller, the prediction begins to underestimate the lattice points, which is entirely expected since $\hat{\rho}_{\text{PS}}(\omega)$ only contains information about the two lowest-lying states $\pi$ and $\pi^{*}$, and for small values of $\tau$ the higher-excited states begin to exert a greater influence on the behaviour of $\hat{C}_{\text{PS}}(\tau)$. Indeed, we observe quantitative agreement for $\tau \gtrsim 0.0007 \, \text{MeV}^{-1} \approx m_{\pi^{*}}^{-1}$, which is the smallest screening length resolved by the spatial correlator. Figure~\ref{T_corr_plot} therefore provides robust evidence that the spectral function in Eq.~\eqref{commutator_PS} is fully consistent with the lattice correlator data from Refs.~\cite{Rohrhofer:2019qwq,Rohrhofer:2019qal} on length scales $\ell \gtrsim m_{\pi^{*}}^{-1}$ at $T=220$ MeV.

\section{Conclusions and outlook} 
\label{concl}

Understanding the non-perturbative structure of correlation functions at finite temperature is essential for obtaining a correct physical interpretation of how particles behave in the presence of a thermal medium. The locality of fields in particular imposes significant constraints, including the existence of a spectral representation for the Euclidean spatial correlator $C(x_{3})$. With this representation we demonstrate that under certain conditions the large-$x_{3}$ behaviour of $C(x_{3})$ is dominated by the lowest-energy particle states, and that in such a regime the spectral function can be directly extracted from $C(x_{3})$ data, avoiding the well-known inverse problem. Applying these results to lattice QCD data for the light-quark spatial pseudo-scalar meson correlator in the temperature range $220-960 \, \text{MeV}$ we are able to determine the explicit contributions to the spectral function $\rho_{\text{PS}}(\omega,\mathbf{p})$ from the lowest-lying states. As a non-trivial test, we demonstrate that $\rho_{\text{PS}}$ is consistent with the corresponding temporal lattice correlator data for $T= 220 \, \text{MeV}$. \\

\noindent
We find that the pion $\pi$ and its first-excited state $\pi^{*}$ dominate the behaviour of $\rho_{\text{PS}}$, and that the $\pi$ is distinguishable in a range of temperatures above the chiral pseudo-critical temperature $T_{\!\text{pc}}$. This suggests that non-perturbative effects continue to play an important role above $T_{\!\text{pc}}$ even for hadronic states composed of light quarks, which is consistent with the expectations based on the approximate realisation of chiral spin symmetry. Although this study focused on the spectral properties of correlators involving pseudo-scalar meson operators, this approach can be generalised to meson states with higher spin and different hadronic states, as well as to regimes with non-vanishing baryon density. This work therefore represents a step towards the analytic non-perturbative description of the QCD phase diagram.

\section*{Acknowledgements}
The work of P.~L. and O.~P.~is supported by the Deutsche Forschungsgemeinschaft (DFG, German Research Foundation) through the Collaborative Research Center CRC-TR 211 ``Strong-interaction matter under extreme conditions'' -- Project No. 315477589-TRR 211. O.~P.~also acknowledges support by the State of Hesse within the Research Cluster ELEMENTS (Project ID 500/10.006).

\bibliography{refs_conf22}

\end{document}